\documentclass[aps,prl,reprint,superscriptaddress,nofootinbib]{revtex4-2}

\usepackage{amsmath,amssymb,bm,mathrsfs}
\usepackage{graphicx}
\usepackage{microtype}
\usepackage{hyperref}
\hypersetup{hidelinks}

\newcommand{\dd}{\,\mathrm d}
\newcommand{\e}{\mathrm e}
\newcommand{\StressV}{\mathbb S}
\newcommand{\HeatV}{\mathbb H}
\newcommand{\Kthree}{\mathbb K}
\newcommand{\cB}{\mathcal B}
\newcommand{\He}{\operatorname{He}}
\newcommand{\Cpec}{\bm C}

\begin{document}

\title{Exact Collision Vertex for Stress and Heat Flux}

\author{Ilya Karlin}
\affiliation{Department of Mechanical and Process Engineering, ETH Zurich, CH-8092 Zurich, Switzerland}

\date{September 15, 2026}

\begin{abstract}
For the hard-sphere Boltzmann equation, we derive an exact nonlinear generating vertex for collisional stress and heat-flux production.  The complete two-input Hermite hierarchy for stress resums to a universal traceless tensor times one scalar radial law, while the heat-flux vertex follows from the same law by an exact contraction.  Taylor differentiation yields universal bilinear collision tensors with exact stress and heat-flux gradings, without a finite moment closure.  DSMC tests, including wall-driven flows and a bimodal Mach-5 shock, recover independently sampled Boltzmann productions and the predicted convergence, while the converged production is independent of the thermal Gaussian foot used for its Hermite representation.
\end{abstract}

\maketitle

The hard-sphere gas occupies a special place in kinetic theory.  The same elementary binary collision rule underlies Newtonian many-particle dynamics, the Boltzmann equation, and stochastic binary-collision descriptions such as Kac--Nanbu processes and direct simulation Monte Carlo (DSMC) \cite{GallagherSaintRaymondTexier2014,Kac1956,Bird1994}.  Yet the hard-sphere collision rate is proportional to relative speed; unlike Maxwell molecules, this velocity dependence destroys finite polynomial closure of the collision moments \cite{ChapmanCowling1970}.

These different descriptions meet directly at \emph{collision production}, where the hard-sphere obstruction becomes concrete.  We set the particle mass $m=1$, let $\Cpec=\bm v-\bm u[f]$ denote the peculiar velocity, and use an overdot for the collision contribution only.  The stress and heat-flux productions of an arbitrary distribution function are
\begin{equation}
\begin{aligned}
 \dot\pi_{\alpha\beta}[f]&=\int C_{\langle\alpha}C_{\beta\rangle}Q(f,f)\,\dd\bm v,\\
 \dot q_{\alpha}[f]&=\frac12\int C^2C_{\alpha}Q(f,f)\,\dd\bm v .
\end{aligned}
\label{eq:classical-productions}
\end{equation}
These are the conventional Boltzmann quantities from which viscosity and thermal conductivity are extracted.
The standard hard-sphere collision integral \cite{Boltzmann1872} and its equivalent Carleman form \cite{Carleman1933} may be written as
\begin{align}
 Q(f,f)(\bm v)
 &=\int\!\dd\bm v_*\!\int_{S^2}\!\dd\bm\omega\,
 B_{\rm HS}\,[f'f_*'-ff_*]\nonumber\\
 &=\kappa\iint\delta(\bm x\!\cdot\!\bm y)
 \left[f_{\bm x}f_{\bm y}-f_{\bm x+\bm y}f_{\bm0}\right]\dd\bm x\dd\bm y.
\label{eq:Q-Carleman-main}
\end{align}
Here $B_{\rm HS}\propto|\bm v-\bm v_*|$ is the hard sphere collision kernel and the second line fixes the normalization $\kappa=\sigma_{\rm HS}^2$ in the diameter convention used below. The Carleman form (second line) enforces orthogonality of auxiliary vectors $\bm x\cdot\bm y=0$ through the corresponding Dirac delta-function and is based on translations $f_{\bm h}=f(\bm v+\bm h)$. Both forms are equivalent.

Choose a normalized Gaussian reference centered at the physical local mean, $\bm U=\bm u[f]$,
\begin{equation}
 M(\bm\xi)=(2\pi)^{-3/2}\e^{-\xi^2/2},\qquad
 \bm\xi=\frac{\bm v-\bm U}{\sqrt\Theta},
\label{eq:reference-M}
\end{equation}
write $f=n\Theta^{-3/2}M\Phi$, and scale out $\Gamma_{\rm HS}=\kappa\sqrt\pi$.  Polarizing the quadratic map gives an exact dimensionless bilinear operator
\begin{equation}
 M\cB(\Phi,\Psi)=\Gamma_{\rm HS}^{-1}Q^{\rm pol}(M\Phi,M\Psi),
\label{eq:B-bridge-main}
\end{equation}
with $$Q^{\rm pol}(f,g)=\frac{1}{2}[Q(f+g,f+g)-Q(f,f)-Q(g,g)].$$  The bilinear production forms are then
\begin{equation}
\begin{aligned}
 \mathcal P_{\alpha\beta}(\Phi,\Psi)
 &=\int \xi_{\langle\alpha}\xi_{\beta\rangle}M\cB(\Phi,\Psi)\dd\bm\xi,\\
 \mathcal Q_{\alpha}(\Phi,\Psi)
 &=\frac12\int(\xi^2-5)\xi_{\alpha} M\cB(\Phi,\Psi)\dd\bm\xi,
\label{eq:bilinear-productions}
\end{aligned}
\end{equation}
so that the relation to the conventional form \eqref{eq:classical-productions} is
\begin{equation}
\begin{aligned}
 \dot\pi_{\alpha\beta}&=\Gamma_{\rm HS}n^2\Theta^{3/2}\mathcal P_{\alpha\beta}(\Phi,\Phi),\\
 \dot q_{\alpha}&=\Gamma_{\rm HS}n^2\Theta^2\mathcal Q_{\alpha}(\Phi,\Phi).
\end{aligned}
\label{eq:physical-rescale}
\end{equation}

Polynomial representations of the collision term are classical.  Grad gave a general Hermite expansion of the binary collision term \cite{Grad1949}, while Burnett--Sonine functions provide the corresponding irreducible angular--radial organization \cite{Burnett1935,ChapmanCowling1970}.  Kumar developed an irreducible-tensor formulation of the nonlinear collision coefficients, and subsequent hard-sphere work made high-order matrix elements systematically accessible through recurrence and Hermite-spectral methods \cite{Kumar1966,Ender2012,WangCai2019}.  

The present question is different: instead of evaluating this infinite array coefficient-by-coefficient or truncating the incoming hierarchy, we fix a physical outgoing observable and ask whether its complete nonlinear two-input row can be summed in closed form.  For stress and heat flux the answer is yes; to our knowledge, such a closed hard-sphere generating law has not been given.
Thus the result is not a finite moment closure but an exact bilinear production law with universal coefficient tensors that can be tabulated once and applied to independently generated distributions.  We test this directly below in wall-driven DSMC and in a stationary Mach-$5$ shock with a strongly bimodal local distribution.

The object to be organized is the complete two-input Hermite row of the bilinear collision map, with both incoming multiindices unrestricted.  We use the Fock-space representation of the Hermite hierarchy developed for collision operators in Refs.~\cite{KarlinMaxwell2026,KarlinLinearHS2026}.  Let $|\bm\nu\rangle$ denote the orthonormal number basis of the standard bosonic Fock space \cite{Fock1932,Berezin1966}, and define the polynomial realization
\begin{equation}
 \mathcal R_P|\bm\nu\rangle
 =\frac{\He_{\bm\nu}(\bm\xi)}{\sqrt{\bm\nu!}},
 \qquad \bm\nu!=\nu_1!\nu_2!\nu_3!.
\label{eq:Fock-Hermite-map}
\end{equation}
Thus $\Phi=\sum_{\bm\nu}h_{\bm\nu}\He_{\bm\nu}/\sqrt{\bm\nu!}$ corresponds exactly to $|\Phi\rangle=\sum_{\bm\nu}h_{\bm\nu}|\bm\nu\rangle$.  No quantum dynamics is implied.  The polarized collision map is represented exactly by
\begin{equation}
 \widehat{\cB}(|\Phi\rangle,|\Psi\rangle)
 =\mathcal R_P^{-1}\cB(\mathcal R_P|\Phi\rangle,\mathcal R_P|\Psi\rangle).
\label{eq:Fock-lift-main}
\end{equation}
For independent complex labels $\bm z,\bm u,\bm w\in\mathbb C^3$, introduce the standard unnormalized coherent kets and analytic coherent bras \cite{Perelomov1986} using canonical ladder operators, $\hat{a}_\alpha$, $\hat{a}_\alpha^\dagger$, $[\hat a_\alpha,\hat a_\beta^\dagger]=\delta_{\alpha\beta}$, $\hat{a}_\alpha |0\rangle=0$:
\begin{equation}
 |\bm z\rangle=\e^{\bm z\cdot\hat{\bm a}^\dagger}|0\rangle
 =\sum_{\bm\nu}\frac{\bm z^{\bm\nu}}{\sqrt{\bm\nu!}}|\bm\nu\rangle,
 \qquad
 \langle\bm w|=\langle0|\e^{\bm w\cdot\hat{\bm a}}.
\label{eq:coherent-abstract}
\end{equation}
Thus $\bm z,\bm u$ generate the two incoming Hermite indices and $\bm w$ the outgoing one.  The three-leg symbol
\begin{equation}
 \Kthree(\bm w;\bm z,\bm u)
 =\langle\bm w|\widehat{\cB}(|\bm z\rangle,|\bm u\rangle)
\label{eq:three-leg-abstract}
\end{equation}
is their generating function: Taylor coefficients at the coherent origin recover the complete collision tensor (Supplemental Material~\cite{SupplementalMaterial}).
In the polynomial realization $\mathcal R_P$, the ladder operators act as
\begin{equation}
 \mathcal R_P\hat a_{\alpha}\mathcal R_P^{-1}=\partial_{\xi_{\alpha}},\qquad
 \mathcal R_P\hat a_{\alpha}^\dagger\mathcal R_P^{-1}
 =\xi_{\alpha}-\partial_{\xi_{\alpha}},
\label{eq:Hermite-intertwining-main}
\end{equation}
and the coherent ket becomes
\begin{equation}
 \mathcal R_P|\bm z\rangle
 \equiv E_{\bm z}(\bm\xi)
 =\e^{\bm z\cdot\bm\xi-z^2/2}
 =\sum_{\bm\nu}\frac{\He_{\bm\nu}(\bm\xi)}{\bm\nu!}\bm z^{\bm\nu}.
\label{eq:coherent-Hermite-main}
\end{equation}
The series fixes our probabilists' Hermite convention.  
The crucial computational observation is the identity
\begin{equation}
 M(\bm\xi)E_{\bm z}(\bm\xi)=M(\bm\xi-\bm z).
\label{eq:shifted-Maxwellian-main}
\end{equation}
For real $\bm z$ this is a displaced Maxwellian; for complex $\bm z$ it is its analytic continuation.  Since $\mathcal R_P$ identifies the Fock inner product with the weighted Hermite inner product in $L^2(M\dd\bm\xi)$, the same abstract symbol is evaluated in velocity space as
\begin{equation}
 \Kthree(\bm w;\bm z,\bm u)
 =\int M(\bm\xi)E_{\bm w}(\bm\xi)\,
 \cB(E_{\bm z},E_{\bm u})(\bm\xi)\dd\bm\xi .
\label{eq:three-leg}
\end{equation}
Substitution of these shifted Gaussians into the polarized Carleman form and Gaussian integration over $\bm\xi$ give the fused kernel.  With $\bm Z=\bm z+\bm u$, $\bm r=\bm z-\bm u$, $\bm a=\bm x+\bm y$, and $\bm b=\bm x-\bm y$,
\begin{align}
 \Kthree(\bm w;\bm Z,\bm r)
 ={}&\frac{\e^{-r^2/4}}{8\pi^2}\iint
 \delta(\bm x\!\cdot\!\bm y)\e^{-a^2/4}
 \e^{(\bm Z-\bm a)\cdot\bm w/2-w^2/4}\nonumber\\
 &\times\left[\cosh\!\left(\frac{\bm r\cdot\bm b}{2}\right)
 -\cosh\!\left(\frac{\bm r\cdot\bm a}{2}\right)\right]\dd\bm x\dd\bm y.
\label{eq:fused-K-main}
\end{align}
Stress and heat-flux production are obtained by the outgoing projections
\begin{equation}
\begin{aligned}
 \StressV_{\alpha\beta}&=\left.\mathscr D^{(2)}_{\alpha\beta}\Kthree\right|_{\bm w=0},&
 \mathscr D^{(2)}_{\alpha\beta}&=\partial_{w_{\alpha}}\partial_{w_{\beta}}-\frac13\delta_{\alpha\beta}\Delta_{\bm w},\\
 \HeatV_{\alpha}&=\left.\mathscr D_{\alpha}^{(q)}\Kthree\right|_{\bm w=0},&
 \mathscr D_{\alpha}^{(q)}&=\frac12\Delta_{\bm w}\partial_{w_{\alpha}}.
\end{aligned}
\label{eq:coherent-projectors}
\end{equation}
We call $\StressV_{\alpha\beta}$ and $\HeatV_{\alpha}$ the stress and heat-flux collision vertices.  Evaluation in closed form begins with stress.
The collision invariants and parity eliminate the common-label contributions from the stress sector.  Rotational covariance then leaves $\bm r=\bm z-\bm u$ as the only available vector, forcing
\begin{equation}
 \StressV_{\alpha\beta}(\bm z,\bm u)=V(r^2)r_{\langle\alpha}r_{\beta\rangle}.
\label{eq:stress-factorized}
\end{equation}
The Supplemental Material performs the intervening Carleman-frame and shifted-Gaussian reductions explicitly.  
For real $s=r^2$ the scalar hard-sphere law is
\begin{equation}
\begin{aligned}
 V(s)={}&-\frac{s^2+4s-12}{2s^2}\exp\left({-\frac s4}\right)\\
 &-\frac{\sqrt\pi}{4}\,
 \frac{s^3+6s^2-12s+24}{s^{5/2}}
 \operatorname{erf}\!\left(\frac{\sqrt s}{2}\right),
\end{aligned}
\label{eq:V-erf}
\end{equation}
with a removable singularity at the origin.  Its Taylor series begins
\begin{equation}
 V(s)=-\frac85-\frac{2}{35}s+\frac{1}{1260}s^2
 -\frac{1}{55440}s^3+\cdots .
\label{eq:V-series}
\end{equation}
The entire complex continuation is given in the Supplemental Material.  

The scalar law \eqref{eq:V-erf} is the main result of this Letter.  Together with the coherent generating identity, it gives the exact hard-sphere stress-production row for every admissible distribution for which the Hermite representation and collision weak form are defined; no finite incoming truncation or closure is involved.

Applying the cubic projector to the same kernel yields the heat flux collision vertex 
\begin{equation}
 \HeatV_{\alpha}(\bm z,\bm u)=\frac12 Z_{\beta}\StressV_{\alpha\beta}(\bm z,\bm u).
\label{eq:heat-stress}
\end{equation}
Thus the exact cubic heat-flux vertex requires no new radial law: it is determined entirely by the same scalar function \eqref{eq:V-erf}. 

The coherent labels $\bm z,\bm u$ now disappear from the physical calculation.  Taylor differentiation at $\bm z=\bm u=0$ defines universal Hermite collision tensors
\begin{equation}
\begin{aligned}
 P_{\alpha\beta;\bm\nu\bm\lambda}
 &=\left.
 \frac{\partial_{\bm z}^{\bm\nu}\partial_{\bm u}^{\bm\lambda}\StressV_{\alpha\beta}}
 {\sqrt{\bm\nu!\bm\lambda!}}\right|_0,\\
 Q_{\alpha;\bm\nu\bm\lambda}
 &=\left.
 \frac{\partial_{\bm z}^{\bm\nu}\partial_{\bm u}^{\bm\lambda}\HeatV_{\alpha}}
 {\sqrt{\bm\nu!\bm\lambda!}}\right|_0.
\end{aligned}
\label{eq:PQ-definition}
\end{equation}
These tensors are precisely the Hermite matrix elements of the bilinear production forms $\mathcal P_{\alpha\beta}$ and $\mathcal Q_{\alpha}$ in Eq.~(\ref{eq:bilinear-productions}); for arbitrary Hermite expansions of the two inputs they are contracted with the corresponding two coefficient sets.  The physical quadratic collision term is obtained by setting $\Psi=\Phi$, which returns Eq.~(\ref{eq:physical-rescale}).
Because $V(r^2)$ is radial, its even derivatives at the origin are fully symmetric isotropic pairing tensors, while odd derivatives vanish.  Exactly two derivatives act on $r_{\langle\alpha}r_{\beta\rangle}$; hence the stress hierarchy has degrees $2,4,6,\ldots$ and is generated by $V^{(m)}(0)$ multiplying a universal Wick/STF pairing tensor.  Heat follows algebraically from Eq.~(\ref{eq:heat-stress}),
\begin{equation}
 Q_{\alpha;\bm\nu\bm\lambda}
 =\frac12\sum_{\beta}\left[
 \sqrt{\nu_{\beta}}\,P_{\alpha\beta;\bm\nu-\bm e_{\beta},\bm\lambda}
 +\sqrt{\lambda_{\beta}}\,P_{\alpha\beta;\bm\nu,\bm\lambda-\bm e_{\beta}}
 \right],
\label{eq:Q-from-P}
\end{equation}
where $\bm e_\beta$ is the unit multi-index in the $\beta$-direction.
Thus, $Q$ has degrees $3,5,7,\ldots$.  Both tensors are symmetric under exchange of the incoming legs $\bm\nu,\bm\lambda$.  For example,
\begin{equation}
\begin{aligned}
 P_{xy;(110),(000)}&=-\frac85,&
 P_{xy;(100),(010)}&=\frac85,\\
 P_{xx;(200),(000)}&=-\frac{16\sqrt2}{15}.&&
\end{aligned}
\label{eq:leading-P}
\end{equation}
The closed pairing formula and scalar coefficient table are given in the Supplemental Material.

This coefficient representation gives a direct interface to an arbitrary measured distribution.  For a DSMC cell, write
\begin{equation}
 \Phi(\bm\xi)=\sum_{\bm\nu}h_{\bm\nu}
 \frac{\He_{\bm\nu}(\bm\xi)}{\sqrt{\bm\nu!}},
\end{equation}
and let the particle velocities $\bm v_p$ sample the coefficients as
\begin{equation}
 h_{\bm\nu}=\frac1{N_p}\sum_{p=1}^{N_p}
 \frac{\He_{\bm\nu}(\bm\xi_p)}{\sqrt{\bm\nu!}},\qquad
 \bm\xi_p=\frac{\bm v_p-\bm U}{\sqrt\Theta}.
\label{eq:particle-hermite}
\end{equation}
The analytic prediction is then
\begin{equation}
\begin{aligned}
 \dot\pi_{\alpha\beta}&=\Gamma_{\rm HS}n^2\Theta^{3/2}
 \sum_{\bm\nu,\bm\lambda}P_{\alpha\beta;\bm\nu\bm\lambda}h_{\bm\nu}h_{\bm\lambda},\\
 \dot q_{\alpha}&=\Gamma_{\rm HS}n^2\Theta^2
 \sum_{\bm\nu,\bm\lambda}Q_{\alpha;\bm\nu\bm\lambda}h_{\bm\nu}h_{\bm\lambda}.
\end{aligned}
\label{eq:physical-PQ}
\end{equation}
No effective parameters $\bm r$ or $\bm Z$ used in the vertex construction are assigned to the gas.  In the numerical tests the sums are evaluated through increasing Hermite order only to demonstrate convergence; no finite-order closure is assumed.  Gaussian-weighted $L^2$ completeness and continuity of the hard-sphere weak forms justify the infinite representation, as detailed in the Supplemental Material.

\begin{figure*}[t]
\centering
\includegraphics[width=0.96\textwidth]{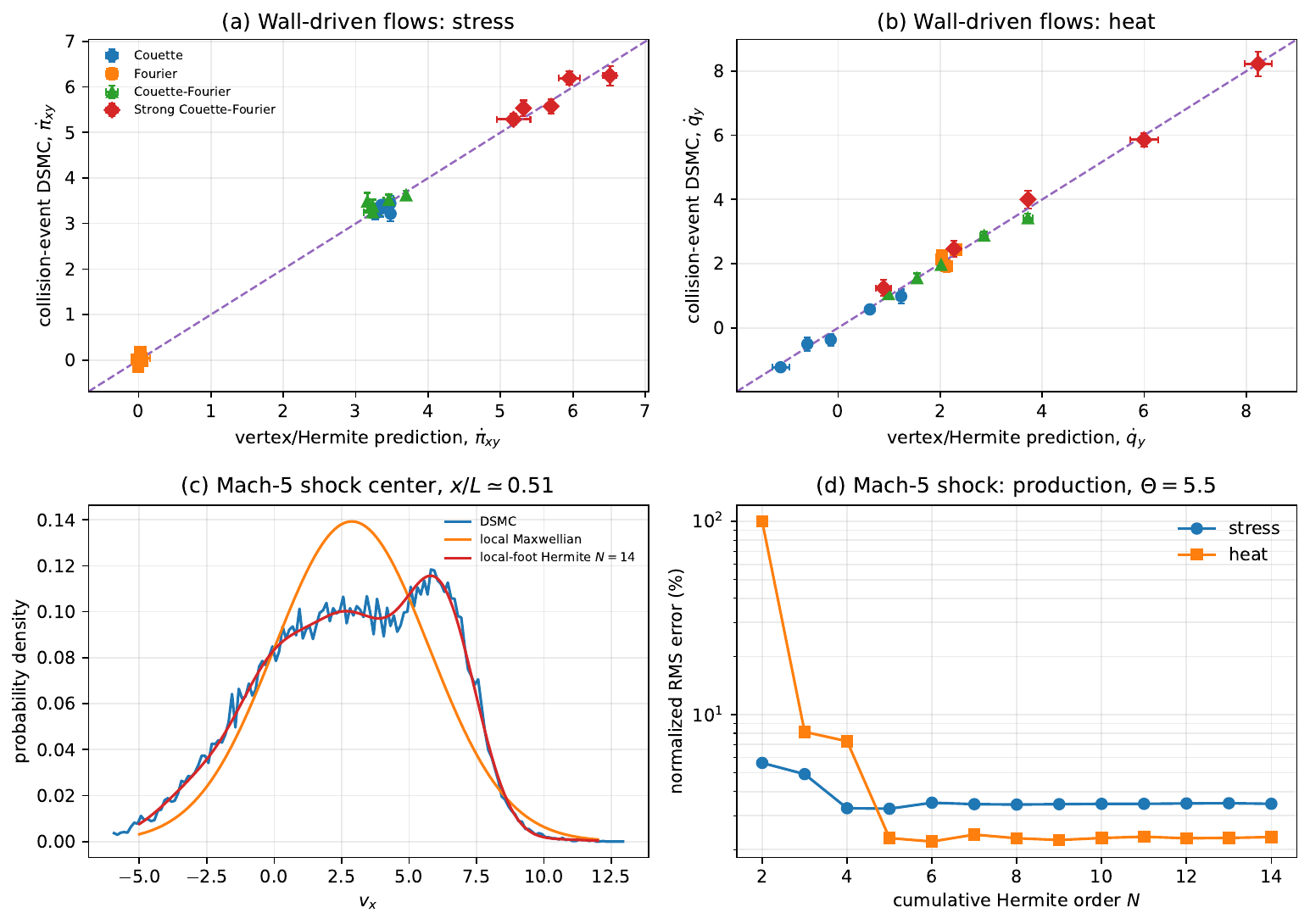}
\caption{Validation on independently generated distributions.  (a),(b) Wall-driven parity plots: $x=$ $N=8$ vertex/Hermite prediction, $y=$ accepted-collision DSMC measurement.  Each symbol is one spatial sample; shape/color identifies the flow.  The dashed diagonal denotes a perfect match.  Error bars show sampling uncertainty.  (c) Mach-5 shock-center distribution, local-Maxwellian Hermite foot, and $N=14$ reconstruction.  (d) Normalized RMS production error versus cumulative Hermite order.}
\label{fig:natural-validation}
\end{figure*}

Figure~\ref{fig:natural-validation} tests this route without preparing coherent states.  Four Couette/Fourier drivings are sampled at five interior positions, including two on the Knudsen-layer side of the flow.  At each position we compare accepted-collision moment changes, direct sampling of the Boltzmann weak form, and Eq.~(\ref{eq:physical-PQ}) from measured Hermite coefficients.  Relative to the weak form, the $N=8$ normalized RMS discrepancies are $1.78\%$ for stress and $2.02\%$ for heat; the near-wall subset gives $1.78\%$ and $1.80\%$.

A more severe test is a stationary Mach-5 shock with a visibly bimodal center distribution [Fig.~\ref{fig:natural-validation}(c)].  With fixed reference $\Theta=5.5$, the $N=14$ RMS discrepancies are $3.45\%$ for normal-stress and $2.32\%$ for heat production.  Quadrupling the sampling statistics lowers these to $1.79\%$ and $0.93\%$ without moving the onset of the plateau, identifying the remainder as predominantly statistical.  Regrouping that reconstruction by the exact total incoming degree gives stress $11.25\%\to1.76\%$ from degree 2 through 4, while heat gives $14.15\%\to1.62\%\to0.89\%$ through degrees 3, 5, and 7.  The different convergence signatures therefore directly reflect the two exact gradings.  Finally, replacing $\Theta=5.5$ at the shock center by the local-Maxwellian reference $T_{\rm loc}\simeq8.20$ changes the converged production by only $0.058\%$ for stress and $0.009\%$ for heat, although that Maxwellian is a poor pointwise approximation to the distribution (Supplemental Material).

For real coherent labels Eq.~(\ref{eq:stress-factorized}) also has a direct particle interpretation.  Since the relative velocity of the two displaced unit-temperature Maxwellians is Gaussian, the stress collision vertex can equivalently be written as an expectation over a random vector with mean $\bm r$ and covariance $2I$,
\begin{equation}
 \StressV_{\alpha\beta}(\bm r)
 =-\frac{\sqrt\pi}{4}
 \mathbb E_{\bm G\sim N(\bm r,2I)}
 \left[|\bm G|G_{\langle\alpha}G_{\beta\rangle}\right].
\label{eq:gaussian-expectation}
\end{equation}
Direct relative-velocity Monte Carlo and collision-level DSMC with prepared displaced-Maxwellian legs verify this identity and the heat--stress contraction (Supplemental Material).  In Fig.~\ref{fig:natural-validation}, by contrast, $\bm r$ and $\bm Z$ remain purely auxiliary generating variables.

We have shown that the nonlinear hard-sphere collision operator has a global organization that is hidden in its infinite Hermite coefficient array.  For a fixed physical outgoing observable, the unrestricted two-input stress row resums before coefficient extraction to one scalar radial law, and heat flux is an exact algebraic descendant of the same vertex.  This differs from a finite moment closure: all incoming Hermite sectors remain available, while the Taylor coefficients of the closed vertex supply their exact grading.  The distinct stress and heat-flux convergence observed through the Mach-5 shock is a direct numerical manifestation of that grading.

The construction also separates physics from representation.  The thermal Gaussian foot, Hermite coefficients, Fock states, and coherent labels are computational coordinates; after convergence they recover the same representation-independent Boltzmann map $f\mapsto(\dot{\bm\pi},\dot{\bm q})$.  The real coherent slice gives an independent particle interpretation through the relative-velocity Gaussian, while higher radial projections remain differential descendants of the same scalar law (Supplemental Material).  For example, their linear limits agree with the companion Fock analysis of the linearized hard-sphere collision integral  \cite{KarlinLinearHS2026} and recover classical ratios of the second to first Sonine polynomial approximation of viscosity and thermal conductivity \cite{ChapmanCowling1970}, $\eta_2/\eta_1=205/202$ and $\lambda_2/\lambda_1=45/44$, respectively.  Beyond the exceptional Maxwell-molecule case, classical hard spheres therefore provide an explicit interaction for which selected nonlinear Boltzmann production hierarchies admit exact analytic resummation.

\subsection*{Data availability} 
\noindent A machine-readable sparse $P,Q$ table of coefficients \eqref{eq:PQ-definition} through total incoming degree 15, its generator and data from simulations are included in the accompanying reproducibility archive [URL to be inserted].

\subsection*{Acknowledgement of AI assistance}
\noindent During the development and preparation of this manuscript, the author used
ChatGPT (OpenAI, GPT-5.6 Sol) as an interactive research and writing assistant.
Its use included discussion and critical examination of mathematical derivations, organization and analysis of simulation data, and development of the presentation.  All results and final text were reviewed and accepted by the
author, who assumes full responsibility for the content of the manuscript.

\nocite{Bargmann1961,Slater1960,NISTHandbook2010}
\bibliographystyle{apsrev4-2}

%

\end{document}